\documentclass{article}[12pt]
\pdfoutput=1

\usepackage{graphicx}

\usepackage[dvipsnames,usenames]{color}

\usepackage{amsmath}                        

\definecolor{DarkGreen}{rgb}{0.5,0.8,0.6}   
\definecolor{RGBblack}{rgb}{0.0,0.0,0.0}    

\definecolor{grau}{rgb}{0.8,0.8,0.8}

\newcommand{\chen}[1]{\color{orange}}

\usepackage[table]{xcolor}
\usepackage[thinlines]{easytable}
\usepackage{multirow,multicol,rotating}
\usepackage{url}
\usepackage{rotating}
\usepackage{amsmath}

\usepackage{graphicx}
\usepackage{amsmath,bm,natbib}                        

\usepackage{longtable}
\usepackage{setspace}

\def\cD{\mathcal{D}}

\begin{document}

\title{  
  A Bayesian Interval Dose-Finding Design Addressing Ockham's
   Razor: mTPI-2}
  
\author{Wentian Guo$^1$, Sue-Jane Wang$^{2,\#}$, Shengjie Yang$^3$, Suiheng
  Lin$^1$, Yuan Ji $^{3,4,*}$\footnote{koaeraser@gmail.com}\\
$^{1}$Department of Biostatistics, School of Public Health\\ Fudan University, P.R. China \\
$^{2}$Office of Biostatistics/Office of Translational Sciences
 \\ $\;\;\;$ Center for Drug Evaluation and Research
 \\ $\;\;\;$  U.S. Food and Drug Administration\\
$^{3}$Program for Computational Genomics \& Medicine
 \\ $\;\;\;$ NorthShore University HealthSystem\\   
$^{4}$Department of Public Health Sciences
 \\ $\;\;\;$ The University of Chicago\\
$^\#$Disclaimer: This article reflects the views of the author
  and should not be construed \\to represent the views or policies of
  the U.S. Food and Drug Administration
           }
           
\maketitle

\clearpage





\begin{abstract}
There has been an increasing interest in
using interval-based Bayesian designs for dose finding, one of which
is the modified toxicity probability interval (mTPI) method. We show
that the decision rules in mTPI correspond to an optimal rule under a formal Bayesian decision
theoretic framework. However, the probability models in mTPI are
   overly   sharpened by the Ockham's razor, which, while in general helps with
parsimonious statistical inference,  leads to suboptimal decisions in
small-sample inference such as dose finding. We
propose a new framework that   blunts   
the Ockham's razor, and
demonstrate the superior performance of the new method, called mTPI-2. An
online web tool is provided for users who can generate the design,
conduct clinical trials, and examine operating
characteristics of the designs through big data and crowd sourcing. 
\end{abstract}


\section{Introduction}
  Often, phase I trials in diseases like cancer, osteoarthritis,
and psoriasis   aim to find the maximum tolerated dose (MTD),
the highest dose with toxicity rate lower than or close to a pre-specified target
level, $p_T$. As in most statistical inference, an estimated MTD is
usually produced to represent    the true and unknown
MTD. However, the \ estimation is always with noise and the
probability of toxicity for the estimated MTD is never exactly the
same as $p_T$. 
For this reason, 
the statistical community has been   considering   interval-based inference to
account for the variabilities in the toxicity estimates. For example,
\citet{cheung2002simple} propose to treat any dose with toxicity
probability in the ``indifference
interval'' $(p_T - \delta, p_T + \delta) $ as an estimated MTD, as
long as a small
$\delta \in (0, 1)$ is agreed upon   at the design stage   by the clinical team. Later, in \citet{ji2007dose,ji2010modified} and \citet{ji2013modified},   the authors further developed 
toxicity probability interval (TPI) and modified TPI (mTPI) methods,
in which they formally proposed a decision theoretic framework linking the
dose-finding decisions of ``Stay'' (S), ``De-escalation'' (D), and
``Escalation'' (E) with the equivalence interval $EI=(p_T - \epsilon_1, p_T +
\epsilon_2)$, over-dosing interval $OI=(p_T+\epsilon_2, 1)$, and
under-dosing interval $UI=(0, p_T-\epsilon_1)$, respectively. For a given
dose $d$,   the authors   calculate $Pr(p_d \in EI \mid data)$, $Pr(p_d \in OI
\mid data)$, and $Pr(p_d \in UI \mid data)$, three posterior probabilities that the toxicity rate $p_d$ 
belongs to each of the three dosing intervals.   The
authors  
associate the dose-finding decisions with these three posterior
probabilities.  
Distinctively,   inference in mTPI   is directly
linked to the posterior probabilities of the three dosing intervals,
which is different from a class of other interval designs \citep{ivanova2007cumulative,oron2011dose,liu2015bayesian}
that use a point
estimate $\hat{p}_d$ and compare $\hat{p}_d$ with three dosing
intervals. That is, these interval designs do not directly calculate
 posterior probabilities of the intervals. They use the intervals as a
 thresholding device where their inference is still based on a point
estimate of $p_d$. 

Interval-based designs, such as mTPI \citep{ji2010modified} are based on
parametric models and use model-based inference for decision
making. In \citet{ji2013modified} and \citet{yang2015integrated} the superiority of
the interval-based designs over the standard rule-based designs, such
as the 3+3 design is established using massive simulations and crowd sourcing. One critical and distinctive feature of mTPI is its
ability to precalculate all the dose   finding decisions    
in
advance, allowing investigators to examine the decisions before the trial 
starts. Therefore, even though a model-based design, mTPI exhibits the same
simplicity and transparency as rule-based methods. 

However, some decision rules in mTPI could be debated in
practice. For example, when the target toxicity probability $p_T=0.3$,
and 3 out   of   6
patients treated at a dose experience dose
limiting toxicity (DLT) events, mTPI would suggest
``S'', stay at the current dose and enroll more patients to be
treated at the dose. Since the empirical rate is 3/6, or 50\%,
practitioners have argued that the decision should be ``D'',
de-escalation instead of ``S''. Another case is when $p_T=0.3$
and 2 out   of   9 patients experience DLT events at a dose, mTPI would suggest ``S''
as well. Investigators could argue that the decision should be
``E'', escalation since the empirical rate is 2/9, or 22\%. 
For this reason, \citet{yang2015integrated} proposed an ad-hoc remedy that allows the
decision rules in the mTPI design to be modified by users. While this feature allows great
flexibility in practice, it lacks solid statistical justification   and
therefore cannot be properly assessed.  

To this end, we propose mTPI-2, an extension of mTPI that solves the
undesirable issue in the current decision under mTPI. We show that
the suboptimal rules listed above are consequences of the Ockham's
razor \citep{jefferys1992ockham}.   The Ockham's razor usually helps Bayesian inference to
automatically achieve parsimony by favoring simpler
models. However, in the case of dose finding with small sample size,
the Ockham's razor is too sharp and must be blunted.   Otherwise,
anti-intuitive decisions, such as those listed above, will be
generated as a consequence of parsimonious inference under the Ockham's
razor. 
In mTPI-2, we provide a new framework to
blunt the Ockham's razor, which leads to an improved decision table. 

The remainder of the paper is organized as follows. Section 2 is
devoted to Ockham's razor and its role in interval-based
designs. Section 3 proposes mTPI-2 as a solution to blunt the Ockham's
razor with a few simple theoretical results. Section 4 examines the
numerical performance of mTPI-2, in comparison to the 
mTPI design using 
crowd sourcing. Section 5 introduces an
online software that implements both methods and Section 6 ends
the manuscript with a discussion.

\section{Ockham's Razor and Interval-Based Designs}
As an accepted principle in science,   the   Ockham's razor states the
principle that an explanation of the facts should be no more
complicated than necessary \citep{thorburn1918myth,jefferys1990bayesian,
good1967bayesian,mackay1992bayesian,jefferys1992ockham}.
 A direct impact of Ockham's
razor is on model selection, which favors ``smaller'' models if data
can be fit similarly well by different models. 

Usually, in model selection one considers multiple models
$\{M_i; i = 1, \ldots, I\}$, and for each model $M_i$, a set of
parameters $\theta_i$. Bayesian inference involving model selection
typically requires a prior $p(M_i)$ for the candidate model  $i$  and
a prior $p(\theta_i \mid M_i)$ for parameters $\theta_i$ that characterize the parameters of
interests in model $M_i$. Formal posterior inference calculates the
posterior probability of the model $p(M_i \mid data)$ and selects
the model with the largest posterior probability. Numerous papers have
shown that the inference based on the posterior probability $p(M_i
\mid data)$ automatically applies the Ockham's razor, in that models
with more parameters and larger parameter space are penalized. 

In general,   the   Ockham's razor helps Bayesian inference by selecting more
parsimonious models. However, in the case of interval-based designs
for dose
finding, such as mTPI, Ockham's razor is too sharp and leads to
practically undesirable decisions. To see this, we first conduct a
quick review of the mTPI design.

The mTPI design considers three intervals that partition the sample space $(0,
1)$ for the probability of
toxicity $p_d$ at a given dose $d$: 
\begin{eqnarray}
M_E: && p_d \in (0, p_T - \epsilon_1) \nonumber \\
M_S: &&p_d \in (p_T - \epsilon_1, p_T + \epsilon_2) \nonumber \\
M_D: &&p_d \in (p_T + \epsilon_2, 1) \label{eq:model}
\end{eqnarray}
The three intervals can be viewed as three models   $M_i$   with index $i \in
\{E, S, D\}$, where the three letters correspond to the dose-finding
decisions if they are selected. For example, when $M_E$ is selected as
the winning model, the corresponding
decision is ``E'', to escalate from the current dose. 
 Typically, $p_T$ ranges from $0.1$ to $0.3$   in   phase I 
 trials,    and
$\epsilon$'s are usually small, say $\le 0.05$. 
In mTPI, the observed data are
integers  $(x_d, n_d)$, where $n_d$ and $x_d$ represent the
numbers of patients treated at dose $d$ and those who have experienced
DLT events, respectively. Given
$p_d$, the probability of toxicity at dose $d$, $x_d \mid p_d \sim
Bin(n_d, p_d)$ a binomial distribution. The mTPI design assumes that
$p_d \sim Beta(1, 1)$, and the
dose-finding decision rule for
dose $d$ is given by 

\begin{equation}
\cD_{\mbox{mTPI}} =  \arg \max_{i \in \{E, S, D\}} UPM(i, d) \label{eq:mTPI-rule}
\end{equation}
where 
\begin{equation}
  UPM(i, d) = \frac{Pr(M_i \mid \{x_d, n_d\})}{S(M_i)} \label{eq:UPM}
\end{equation}
 is the posterior probability of the interval $M_i$ divided by the
 length of the interval. 

We first show that the decision rule $\cD_{\mbox{mTPI}}$ is optimal if
intervals $M_i$ are considered part of the candidate models in a
model-selection framework. To see this, we introduce an additional
parameter $m_d \in \{M_E, M_S, M_D\}$, which denotes the indicator of
the three candidate models (intervals) to which $p_d$ belongs. In particular, Theorem 1 below shows that decision
 $\cD_{\mbox{mTPI}}$ corresponds to the Bayes rule, the optimal
 decision rule that minimizes the posterior expected loss under a 0-1
 loss function $\ell(a, m_d)$ \citep{berger19880}, defined by 
\begin{equation}
\ell(a=i, m_d=M_j) = \left\{\begin{array}{cc}1, & \mbox{ if } i \ne j; \\ 
                                                0, & \mbox{ if } i = j,
                                                
                \end{array}\right.
\quad \mbox{ for } i, j \in \{E, S, D\}.
\label{eq:0-1loss-1}
\end{equation}
The loss function $\ell(a, m_d)$ states that the loss for taking
action $i$ is 0 if model $M_i$ is the winning model, and 1
otherwise. 

\vskip 3ex

\noindent {\bf Theorem 1.} {\it Given the sampling model $x_d \mid p_d \sim
Bin(n_d, p_d)$ and priors 
\begin{eqnarray*}
p_d \mid m_d = M_i &\sim& \frac{1}{S(M_i)} I(p_d \in M_i) \\
p(m_d = M_i) &=& \frac{1}{3}
\end{eqnarray*}
independently for all doses,
and given the 0-1 loss function $\ell(i, M_j)$ in \eqref{eq:0-1loss-1}
for three decisions, where 
$i,j \in \{E, S, D\}$, decision rule $\cD_{\mbox{mTPI}}$ in
\eqref{eq:mTPI-rule} is
optimal in the sense that it minimizes the posterior expected loss. }

\noindent Proof is given in the Appendix A.
\vskip 3ex

\noindent The Bayes rule $\cD_{\mbox{mTPI}}$ selects the action   $i
\in \{E, S, D \}$   corresponding to
the model $M_i$ with the largest posterior probability. This inference
is subject to Ockham's razor. 
As an example, when $x_d =3$ and $n_d =6$, i.e., the
decision rule $\cD_{\mbox{mTPI}}$ boils down to comparing the $UPM(S,
d)$ and $UPM(D, d)$, which involves the
calculation of the posterior probability $Pr(M_i \mid {x_d, n_d})$ for
$M_S=(p_T - \epsilon_1, p_T + \epsilon_2)$ and $M_D = (p_T +
\epsilon_2, 1)$.   For each model, the size of the model is the
length of the interval in the model.   The model size $S(M_D) = (1-p_T -\epsilon_2)$ is
usually larger than the size $S(M_S) = (\epsilon_1 + \epsilon_2)$
since usually $p_T$ is close to 0.3 or 0.16, and $\epsilon_{1, 2} \le
0.05.$. The posterior probability $Pr(M_i \mid {x_d, n_d})$ can be
written as a difference of incomplete beta functions evaluated at the
boundaries of the two models.   Some theoretical discussion of how
$Pr(M_i \mid x_d, n_d)$ depends on $x_d$, $n_d$ and interval
definitions are given in Appendix B.   
When $x_d=3$ and $n_d=6$, it can be shown that the $UPM(S,
d)$ is larger
than $UPM(D, d)$ for $p_T=0.3$ and
$\epsilon_1=\epsilon_2=0.05$. Consequently, even though the empirical rate $x_d/n_d = 0.5$ is
greater than $p_T=0.3$, mTPI still prefers $S$, to stay at the
current dose. In summary, due to the Ockham's razor   which    prefers more parsimonious
model, in this case model $M_S$ with a shorter interval length, mTPI chooses to
stay at dose $d$ when $x_d=3$ and $n_d=6$. Theoretically, the
exact proof depends on the convexity of the incomplete beta function, which is still an open question \citep{swaminathan2007convexity}
 with no conclusion. Instead, we provide a
numerical illustration next.

\begin{figure}
\begin{center}
  \includegraphics[scale=0.75]{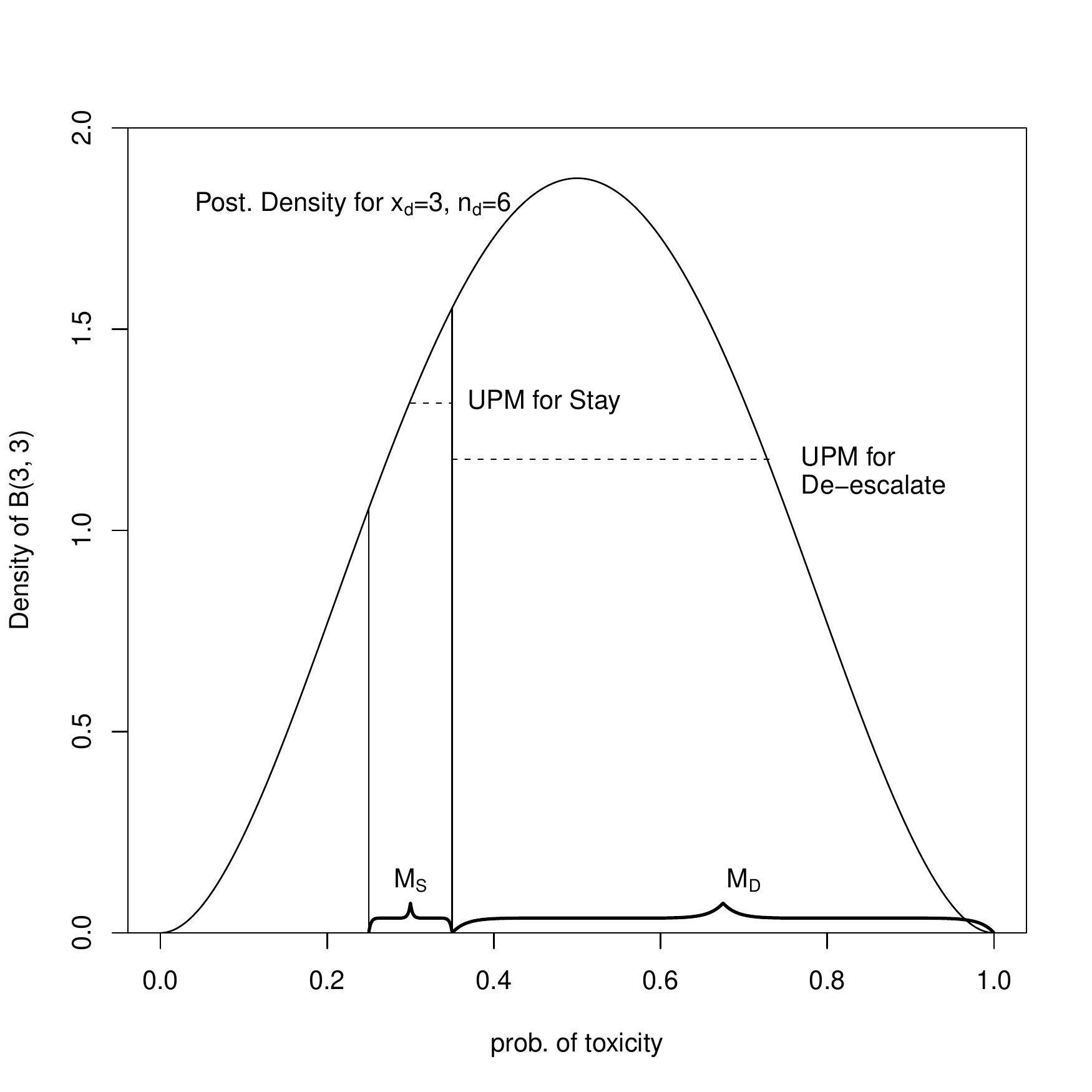}
\caption{An example demonstrating the effect of the Ockham's
  razor in mTPI. Shown is the posterior density of $p_d$ when $x_d=3$ and
  $n_d=6$. Even though the shape of the density suggests that dose $d$
  might be above the MTD, e.g., the posterior mode is to the right of the
  equivalence interval (shown as the two vertical bars), the UPM for
  decision $S$ (stay) is still larger than that the UPM for decision $D$
  (de-escalate). Therefore, mTPI would still choose to ``Stay''
  despite that the shape of the posterior density of $p_d$ indicates otherwise. This is due to the larger size (longer length) of the
  interval $M_D$ than $M_S$ and the Ockham's razor, which prefers the smaller
  model $M_S$.} \label{fig:razor-examp}
\end{center}
\end{figure}

As an example that shows the effect of the Ockham's razor, in Figure \ref{fig:razor-examp}, mTPI will select decision
``S'' even when $x_d=3$ out of $n_d=6$ patients experience the DLT
events, and the posterior distribution is clearly peaked inside the
interval $M_D$. 

\section{A Solution to Blunt the Ockham's Razor: mTPI-2}
\subsection{Decision theoretic framework}
We provide a solution to blunt the Ockham's razor for mTPI and avoid
the undesirable decisions, such as $S$ when 3 out of 6 patients
experience DLT at a given dose. Statistically speaking, there is
nothing wrong with the current decision in mTPI as the Bayesian inference takes into
account the model complexity when choosing the optimal
decision. However, for human clinical trials 
patient safety often outweighs statistical optimality. To this end,
we modify the decision theoretic framework and blunt the Ockham's
razor. 

We call the new class of designs mTPI-2, since the framework is
motivated by that in mTPI. We show next that the framework blunt the Ockham's razor and
leads to safer and more desirable decision rules. Importantly, mTPI-2
preserves the same simple and transparent nature exhibited in mTPI, 
facilitating its practical implementation by both statisticians and
clinicians. 

\begin{figure}
\begin{center}
\includegraphics[scale=0.75]{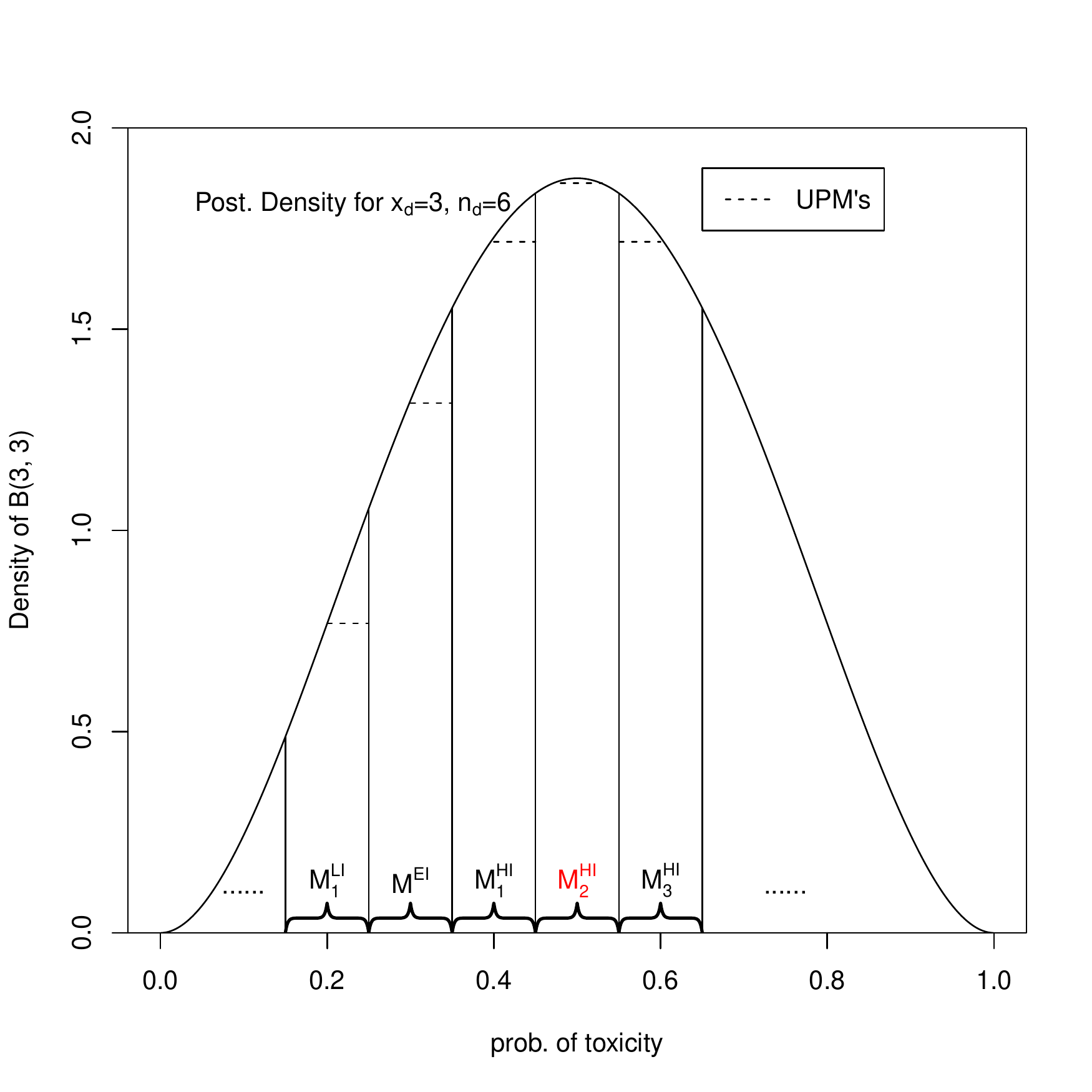}
\caption{An example demonstrating the new framework of mTPI-2. Here,
  $EI$ is the equivalence
 interval $(p_T - \epsilon_1, p_T+\epsilon_2)$, and $LI$ denotes the intervals
 below $EI$, and $HI$ denotes the intervals above $EI$. Interval $M_2^{HI}$
 exhibits the largest UPM and therefore the decision is now $D$, to
 de-escalate.} \label{fig:mTPI-2}
\end{center}
\end{figure}

The basic idea is to divide the unit interval $(0, 1)$ into
subintervals with equal length, given by $(\epsilon_1 +
\epsilon_2)$. This results in multiple intervals with the same length,
which are considered multiple equal-sized models. See Figure \ref{fig:mTPI-2}. For
 clarity, we now denote $EI$ the equivalence
 interval $(p_T - \epsilon_1, p_T+\epsilon_2)$, and $LI$ a set of intervals
 below $EI$, and $HI$ a set of  intervals above $EI$. 
For example, when $p_T=0.3$ and
$\epsilon_1=\epsilon_2=0.05$, the equivalence interval is $EI = (0.25, 0.35)$,  the $LI$ intervals are
$$LI=\{M_1^{LI}=(0.15,
0.25) , \; M_2^{LI}=(0.05, 0.15), \;  M_3^{LI}=(0, 0.05)\}, $$ and the $HI$ intervals are 
\begin{eqnarray*}
HI & = & \{M_1^{HI}=(0.35, 0.45), \; 
M_2^{HI}=(0.45, 0.55), \; M_3^{HI}=(0.55, 0.65), \; M_4^{HI}=(0.65, 0.75),
\;\\ && M_5^{HI}=(0.75, 0.85), M_6^{HI}=(0.85, 0.95), \;
                                M_7^{HI}=(0.95, 1)\}.
\end{eqnarray*} 
The same as mTPI, if the equivalence
 interval $M^{EI}= (p_T - \epsilon_1, p_T + \epsilon_2)$ has the
 largest UPM, it is selected as the winning model and the
 dose-finding decision of mTPI-2 is $S$, stay. If any interval
 $M_i^{HI}$ or $M_i^{LI}$ has the largest UPM, it will be selected as
 the winning model and the dose-finding decision is $D$ or $E$,
 respectively. In Figure \ref{fig:mTPI-2}, for the same posterior
 density corresponding to $x_d=3$ and $n_d=6$, interval $M_2^{HI}$
 exhibits the largest UPM and therefore the decision is now $D$. Note that the same decision theoretic framework as mTPI is in place 
 except that now there are multiple intervals corresponding to $D$ or
 $E$, and the intervals all have the same length, thereby 
 blunting the Ockham's razor. 

\subsection{Optimal rule for mTPI-2}
We again consider a 0-1 loss function $l(a,   m_d   )$, but with multiple
intervals, and multiple decisions. Shown in Table \ref{tab:loss} the
loss function divides the parameter space $(0,1)$ of $p_d$ into $(k_1
+ k_2 +3)$ intervals, with $(k_1 + 1)$ intervals below the equivalence
interval $M^{EI}$ and $(k_2+1)$ intervals above $M^{EI}$. Except  
for    the
two boundary intervals $M_{k_1+1}^{LI}$ and $M_{k_2+1}^{HI}$, all the
intervals have the same length $\delta=(\epsilon_1 +
\epsilon_2)$. The loss
$l(a,   m_d   )$ is a function of action $a$ that selects any of the $(k_1 +
k_2 + 3)$ intervals as the winning model,  and the
parameter $m_d$   indexes the model, which takes one of the intervals $M_i$.  
 
There are a total of $(k_1+k_2+3)$ intervals. Consider the statistical
decision $a$ to select one interval as the winning interval into which the
toxicity probability $p_d$ falls. 
However, selecting a winning 
interval 
must be translated into dose-finding decisions. To this end, we
consider a deterministic mapping. Define $a^* \in \{E, S, D\}$ the
three dose-finding decisions for the trial. Based on ethical
consideration, whenever the statistical decision $a$ is in set $LI$,
$EI$, or $HI$, the corresponding trial decision $a^*$ takes value $E$,
$S$, or $D$, respectively. Mathematically, this means that 
\begin{equation}
a^* = \left\{\begin{array}{ll} 
              E, & \mbox{ if } a \in LI \\
              S, & \mbox{ if } a = EI \\
              D, & \mbox{ if } a \in HI .
\end{array}
\right. \label{eq:a-star}
\end{equation}
The goal is to optimally select $a$, which leads to $a^*$.




\begin{table}[htbp]
\begin{center}
\caption{A loss function of dose   finding    decisions $a$ and model
  parameter $m_d$. Columns are the sample space of $m_d$, i.e., the
  candidate models are the toxicity probability intervals and rows are
  the action values for $a$ and $a^*$ \eqref{eq:a-star}. } \label{tab:loss}
\resizebox{\textwidth}{!}{
\begin{tabular}{|l||c|c|c|c|c|c|c|} \hline
\multicolumn{8}{|c|}{Loss function $\ell(a,m_d)$, for $a$ to select a
  model $\in \{LI, EI, HI \}$ and
  $m_d $ also takes an interval value $\in \{LI, EI, HI\}$.} \\ \hline
& \multicolumn{3}{|c|}{$m_d \in LI$: Intervals below the Equiv. Interval} & {
  $m_d=EI$: Equiv. Interval} & \multicolumn{3}{|c|}{$m_d \in HI$: Intervals above Equiv. Interval} \\ \hline
 Actions $a$, $a^*$ & $M_{k_1+1}^{LI}=(0, p_T - \epsilon_1 - k_1 \delta)$ &  $\cdots$ & $M_{1}^{LI}  = (p_T - \epsilon_1 - \delta,
p_T - \epsilon_1)$ & $M^{EI} = (p_T - \epsilon_1, p_T + \epsilon_2)$ & $M_1^{HI}=(p_T +
\epsilon_2, p_T + \epsilon_2 + \delta)$ & $\cdots$ & $M_{k_2+1}^{HI}=(p_T +
\epsilon_2 + k_2 \delta, 1)$ \\ [2ex] \hline
$a=M_1^{LI}, a^*=E$ & \cellcolor{Gray} 0 & 1 & 1& 1 & 1 &$1$ & $1$ \\ [2ex] \hline
\multicolumn{8}{|c|}{$\cdots$$\cdots$}\\ [3ex]\hline
$a=M_{k_1+1}^{LI}, a^*=E$ & $1$ & $1$ &\cellcolor{Gray} 0 & 1 & $1$ & $1$ & $1$\\ [2ex] \hline
$a=M^{EI}, a^*=S$& $1$ & $1$ & $1$ & \cellcolor{Gray} 0  &  1& $\cdots$ & 1 \\ [2ex] \hline
$a=M_1^{HI}, a^*=D$ & 1 & 1 & 1& 1 & \cellcolor{Gray} 0 &$1$ & $1$ \\ [2ex] \hline
\multicolumn{8}{|c|}{$\cdots$$\cdots$}\\ [3ex] \hline
$a=M_{k_2+1}^{HI}, a^*=D$ & $1$ & $1$ & 1 & 1 & $1$ & $\cdots$ & \cellcolor{Gray} 0 \\ [2ex] \hline
\end{tabular}
}
\end{center}
\end{table}

Assume   that   given $n_d$, $x_d$ follows a binomial distribution,
i.e., $f(x_d \mid n_d, p_d) \propto p_d^{x_d} (1-p_d)^{n_d -
  x_d}$. For $p_d$, 
given interval (model) $m_d=M_i$, assume a prior
\begin{equation}
 p_d \mid m_d=M_i \sim Beta(1, 1) I(p_d \in M_i) \label{eq:prior}.
\end{equation}
Assume
prior probability $p(m_d = M_i)$ is the same for all the models
(intervals), where $M_i \in \cup\{LI, EI, HI\}$. Theorem 2 below
provides the optimal decision rule for mTPI-2.

\bigskip

\noindent {\bf Theorem 2.} {\it The new Bayes rule $\cD_{\mbox{mTPI-2}} \equiv
 \cD_{a^*}$ that
takes action $a^* \in \{E, S, D\}$ corresponds to the Bayes rule $\cD_{a}$
that takes actions $a \in \{LI, EI, HI\}$. Under $\ell(a, m_d)$ in Table
\ref{tab:loss} and the
hierarchical model $\left\{f(x_d \mid n_d, p_d), f(p_d \mid m_d),
  p(m_d)\right\}$ above,
$\cD_{\mbox{mTPI-2}}$ 
is given by the following rule:
\begin{itemize}
 \item If $M_{max} \equiv \arg\max_i Pr(m_d = M_i \mid \{x_d, n_d\}) = EI$,
   $\cD_{\mbox{mTPI-2}} = S$, to Stay.
 \item If $M_{max} \equiv\arg\max_i Pr(m_d = M_i \mid \{x_d, n_d\}) \in LI$,
   $\cD_{\mbox{mTPI-2}} = E$, to Escalate.
 \item If $M_{max} \equiv\arg\max_i Pr(m_d = M_i \mid \{x_d, n_d\}) \in HI$,
   $\cD_{\mbox{mTPI-2}} = D$, to De-escalate.
\end{itemize}}

\noindent Proof is immediate given the fact that $\cD_{a}$ is the
Bayes rule for the loss function in Table \ref{tab:loss} and the
definition in \eqref{eq:a-star}. 

\bigskip

Theorem 2 states that the optimal rule is to first find the interval $M_{max}$
with the largest posterior probability. If $M_{max}$ is the $EI$,
the equivalence interval, stay at the current dose and treat the next
cohort of 
patients at that dose; if $M_{max}$ is   one of the intervals   in $LI$, escalate to and treat
the next cohort of patients at the next higher dose;
if $M_{max}$ is   one of the intervals   in $HI$, de-escalate to and treat the next cohort of
patients at the next lower dose. This
decision rule minimizes the Bayes risk, i.e., the posterior expected
loss. 

\bigskip

\noindent {\bf Corollary 1:} The optimal decision $\cD_{mTPI-2}$ is
equivalent to the following procedure: Assume dose $d$ is the current
dose being used for treatment. 
\begin{enumerate}
 \item Compute $UPM(i, d)$ in \eqref{eq:UPM} for each interval $M_i \in \cup \{LI, EI,
   HI\}$. Let $M_{max}$ be the interval with the largest $UPM$. 
 \item If $M_{max}$ is the $EI$, in $LI$, or in $HI$,  the optimal
   rule $\cD_{mTPI-2}$ is to Stay, Escalate,
   or De-escalate, respectively.
\end{enumerate}

Proof: It suffices to prove $Pr(m_d = M_i \mid \{x_d, n_d\}) = UPM(i,
d),$  which is immediate.

\subsection{Design Algorithm}
The implementation of the mTPI-2 design is as simple and transparent
as mTPI. A decision table of all the optimal decisions in Corollary 1
can be precalculated. See Figure \ref{fig:tables} as an example for a
trial with $p_T=0.3$ and $\epsilon_1=\epsilon_2=0.05$. The
table in Figure \ref{fig:tables}(a) guides all the dose assignment decisions throughout the
trial. For example, suppose a trial has five candidate doses, and dose
3 is being used to treat patients. Then the possible doses for
treating future patients are doses 2, 3, and 4. Record $n_3$ and $x_3$
as the number of patients treated and number of patients experienced
DLT at dose 3, then go to the table entry corresponds to row $x_3$ and
column 
$n_3$, and treat the next cohort of patients  based on the decision in
the table. For example, if $x_3=3$ and $n_3=6$, the decision is $D$ in
Figure \ref{fig:tables}(a), and the next patients will be treated at
dose 2. Note that in contrast, Figure \ref{fig:tables}(a) would
suggest $S$ under mTPI, a now suboptimal decision under mTPI-2.  
More discussion about Figure \ref{fig:tables} will follow next.   The
full algorithm of mTPI-2 is given below, assuming patients are
enrolled in cohorts of size $\ge 1.$ 

\begin{center}
\fbox{\fbox{\parbox{6.5 in}{
\begin{description}%
  \item[] {\bf Optimal decision rule:} Suppose that the current dose is $d$, 
  $d \in \{1,\cdots,D\}$ candidate doses. After the toxicity outcomes
  of the   most recent patient cohort are   
  observed, denote $(x_d, n_d)$ the   current observed   trial data. Select the dose for treating the next cohort among $\{(d-1), d, (d+1)\}$ based on the
  optimal rule $\cD_{mTPI-2}$ in Corollary 1. There are two exceptions: if $d=1$, the next available doses are $\{d, (d+1)\}$; if $d=D$, the next available doses are $\{(D-1), D\}$.%
  \item[] {\bf Trial stopping rule:} Assume $n_1 > 0$. If $Pr(p_1 > p_T \mid
 {x_d, n_d}) > \xi, $ for a large probability $\xi$, say $0.95$,  terminate
 the trial due to excessive toxicity. Otherwise, terminate the trial
 when the maximum sample size is reached. In the special case of
 cohorts of size 1, do not apply the stopping rule $Pr(p_1 > p_T \mid
 {x_d, n_d}) > \xi, $ until three 
 or more patients have been evaluated at a dose. 
  \item[] {\bf MTD selection:}  At the end of the trial, select the dose as the estimated MTD with the smallest
  difference $|\hat{p}^*_d - p_T|$ among all the doses $d$ for
  which $n_d > 0$ and $Pr(p_d > p_T |
 {x_d, n_d}) < \xi$. Here $\hat{p}^*_d$ is the isotonically
 transformed posterior mean of $p_d$, the same as that in the mTPI
 design \citep{ji2010modified}. If two or more doses tie
    for the smallest difference, perform the following rule. Let $p^*$
  denote the transformed posterior mean $\hat{p}_d^*$ of the tied doses. 
\begin{itemize}
  \item If $p^* < p_T$, choose the highest dose among the tied doses.
  \item If $p^* > p_T$, choose the lowest dose among the tied doses.
\end{itemize}
\end{description}%
}}}
\end{center}

\section{ Results}
\subsection{Decision Tables With Bayes Factors}
\begin{figure}[htbp]
\begin{center}
\begin{tabular}{cc}
  \multicolumn{2}{c}{ \hskip -0.3in \includegraphics[scale=0.55,angle=0]{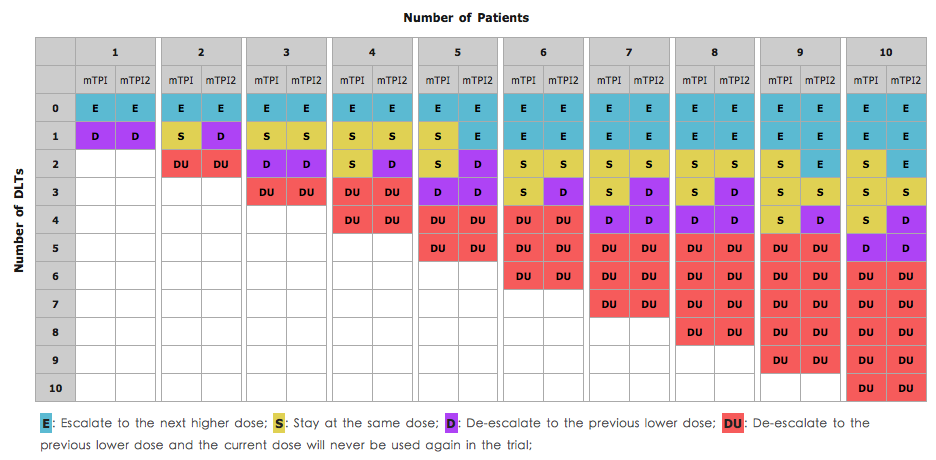}}
 \\
 \multicolumn{2}{c}{(a) A combined decision table for mTPI and mTPI-2. }
  \\ \hline
\hskip -0.5in \includegraphics[scale=0.20]{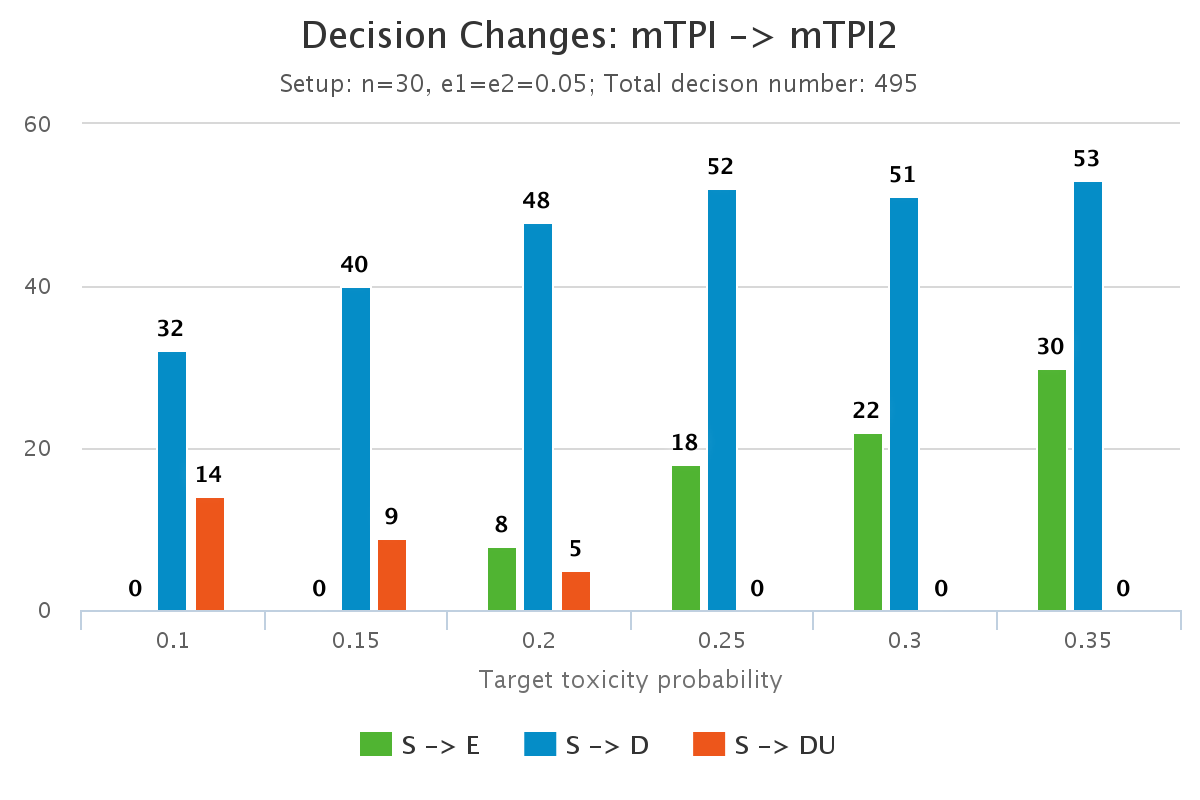}
& \hskip -.3in \includegraphics[scale=0.38]{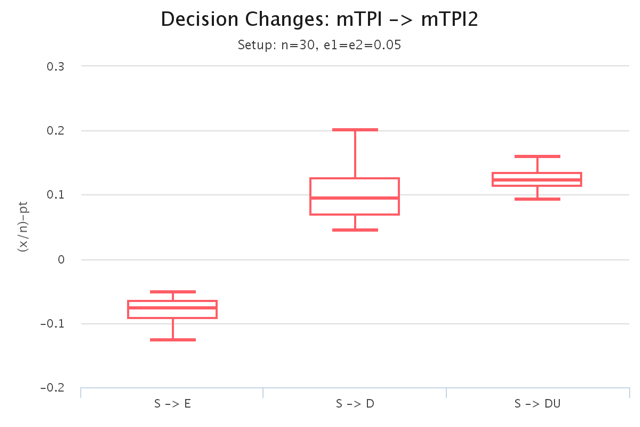} \\
\hskip -.3in {\small (b) Changes between mTPI and
  mTPI-2 for various $p_T$ values.} & {\small (c) A box-plot of $x_d/n_d -
                                      p_T$ values for the changes.}
\end{tabular}
\caption{An example of the optimal decision tables for mTPI and
  mTPI-2. (a) presents 
  decisions for both mTPI and mTPI2. For each ``Number of Patients''
  (column), there are two subcolumns listing the decisions of mTPI and
  mTPI2 side by side. 
  Here, the
  target toxicity probability $p_T=0.30$ and
  $\epsilon_1=\epsilon_2=0.05.$  (b) summarizes the differences in
  decisions between mTPI and mTPI2 with breakdowns of 
  different $p_T$ values. For example, the blue bar denotes a change
  from decision $S$ in mTPI to decision $E$ in mTPI-2. (c) Boxplots of
$(x_d/n_d - p_T)$ for the decisions that are changed in mTPI. The plots show
that when $x_d/n_d < p_T$, decisions $S$ are changed to $E$;
when $x_d/n_d > p_T$, decisions $S$ are changed to $D$ or $DU$. } \label{fig:tables}
\end{center}
\end{figure}

As an interval design, both mTPI and mTPI-2 generate a set of
decisions based on the input values $p_T$, $\epsilon_1$, and
$\epsilon_2$ from physicians. They are summarized in a tabular format,
e.g., those in Figure \ref{fig:tables}. Together, three values define the
equivalence interval $(p_T - \epsilon_1, p_T+\epsilon_2)$ where any
dose with a toxicity probability falling into the interval can be
considered as an MTD. Doses with toxicity probabilities outside the
interval are considered either too low or too high. In a  
dose-finding trial aiming at identifying the MTD,   the decision table can be precalculated for any
values of $p_T \in (0, 1)$ and   $\epsilon_1, \epsilon_2  \ll p_T$,
  and a
sample size which determines column number of the table. Suppose
a sample size $maxN$ is decided for the trial. For each
enumerated integer pairs, $(x, n)$, $0 \le x \le n \le maxN$, the 
decision $\cD_{\mbox{mTPI-2}} \in \{D, S, E\}$ is precalculated. 

  Figures \ref{fig:tables} (a) shows an example of the decision
tables under both designs for $p_T=0.3$ and a sample size of 12. As can
be seen, 
the main improvement of the mTPI-2 design over mTPI is the precise and
``faithful'' decisions that reflect physicians input. For example,
unlike mTPI where a decision $S$ is given when $x_d=3$ toxicity events
are observed
out of $n_d=6$ patients, mTPI-2 recommends $D$, to
de-escalate. Similarly, when $x_d=2$ and $n_d=9$, the decision becomes
$E$ for mTPI-2 instead of $S$ for mTPI. In
essence, mTPI-2 becomes a more ``nimble'' design
due to the effort in blunting the Ockham's razor. 
 Specifically, mTPI 
favors the $EI$ and the decision $S$, to stay, simply because the
equivalence interval has the shortest length and is preferred in the
Bayesian inference due to the Ockham's razor. In contrast, mTPI-2
avoids the Ockham's razor by having equal-lengthed
intervals. Therefore, in
  Figures \ref{fig:tables}(a) the mTPI-2 design shows fewer $S$, more
$D$'s and $E$'s. 

Figure \ref{fig:tables}(b) shows the distribution of
different decisions between mTPI-2 and mTPI for different $p_T$ values
and a large sample size of 30. As can be seen, all the differences are
related to changing the decision $S$ in mTPI to not $S$ ($D$, $E$, or
$DU$) in mTPI-2. In general, many $S$ decisions are changed to $D$ or $E$,
corresponding to the green and blue bars, respectively. Also, when
$p_T < 0.2$, there are no green bars (hence no change from $S$ to $E$), which seems to be sensible since
escalation is less likely when $p_T < 0.2$. In addition, when $p_T \le 0.2$, some $S$
decisions are changed to $DU$ (red bars). That is, some ``stay''
decisions in mTPI are changed to a composite decision in mTPI-2, which
says that first, ``De-escalate'' and second, the current dose
is deemed too toxic and will be removed from the trial. This is a
major modification on the dosing decision. 

We look into why there is
such a big change. For
example, such a change occurs when $p_T=0.1$ and $x_d=3$ out of $n_d=12$
patients experience DLT. Under mTPI, the three intervals are $(0,
0.05)$, $(0.05, 0.15)$, and $(0.15, 1)$. Intuitively, the empirical
toxicity rate equals $x_d/n_d = 0.25$, which is much higher than
$p_T=0.1$. So $D$, de-escalate, should be preferred. However, based on
mTPI the
UPM for $S$ is the largest. The main reason is that the posterior
distribution of $p_d$ is Beta(4, 10) given data $(x_d=3, n_d=12)$,
which has a very light right tail and
puts tiny 
probability mass when $p_d > 0.7$. This allows
Ockham's razor to sharply penalize the right interval $(0.15, 1)$,
which is of length $0.85$.   In contrast, the EI $(0.05, 0.15)$ only
has a length of $0.15$.   As a consequence, the UPM value for each of
the three intervals, defined as
the ratio of interval's posterior probability mass and interval length, favors
the shorter interval $(0.05, 0.15)$ instead of $(0.15, 1)$,  even
though the posterior distribution puts most mass above 0.15. Therefore,
mTPI gives an $S$ for $(x_d=3, n_d=12)$. However, the mTPI-2
design blunts the Ockham's razor and uses sub-intervals with
equal length. Based on the new statistical framework under mTPI-2, the winning subinterval is $(0.25, 0.35)$ and
the optimal decision is $D$. In addition, under mTPI-2 the safety rule is
invoked and therefore $U$ is added. In the case of mTPI, since the
decision is $S$, the safety rule is not even evaluated (mTPI does not
evaluate the safety rule unless the decision is $D$). For these
reasons, when $x_d=3$ and $n_d=12$ at a given dose $d$, mTPI would
stay ($S$) and mTPI would de-escalate and remove dose $d$ from the
trial (due to high toxicity). This example
shows that mTPI-2 is a safer design than mTPI. 

In Figure
\ref{fig:tables}(c) we show that the changes from mTPI decisions to
mTPI-2 decisions are all compatible with the empirical toxicity
rate $x_d/n_d$. That is, mTPI-2 would only change $S$ to $E$ when the empirical
rate is lower than $p_T$, and $S$ to $D$ when the empirical rate is
higher than $p_T$.   

Due to the principled decision-theoretic framework, mTPI-2 calculates
the posterior probability $Pr(m_d = M_i \mid \{x_d, n_d\})$ for each
of the intervals, $M_i \in \{LI, EI, HI\}$. Naturally, the Bayes
factor (BF) between any two intervals can be calculated as 
$$
\mbox{BF}_{ij} = \frac{Pr(m_d = M_i \mid \{x_d, n_d\})}{Pr(m_d = M_j \mid \{x_d, n_d\})},
$$
assuming equal prior probability for each model $M_i$. A value close to 1 means there is only weak evidence supporting one model
or the other. In mTPI-2, in addition to provide
the winning decision in the table, we also display the BF of the
winning decision versus the decision with the second largest posterior
probability. Therefore, all those BF's are greater than 1 but a value
close to 1, say $<1.05$ indicates uncertainty in the decision. Due to
small sample sizes for phase I trials, such weak decisions are not
uncommon as can be seen in Table \ref{tab:BF} below. 

\begin{table}[htbp]
\caption{Decisions in mTPI-2 along with Bayes' factors. For any
  decision that is not ``U'', a Bayes factor (BF) is provided comparing the
  winning decision and the second most likely decision. The BF value
  here 
  is always greater than 1 since   we calculated the BF of the winning
  decision versus the second best decision.   A BF 
  value closer to 1 indicates weaker evidence supporting the winning
  decision.}\label{tab:BF}
\begin{center}
\begin{tabular}{|l|l|lr|lr|lr|lr|}\hline
&& \multicolumn{8}{c|}{Number of Patients} \\ \cline{3-10}
   &  &  3 & (BF) & 6 & (BF) & 9 & (BF) & 12 & (BF) \\[2pt]  \hline 
\multirow{13}{*}{\begin{turn}{-270}{Number of DLTs}\end{turn} }& 0 & E
      & (2.12) & E &(4.47) & E & (9.38) & E &(19.56) \\
& 1 & S & (1.02) & E & (1.29) & E & (2.34) & E & (4.8) \\
& 2 & D & (2.32) & S & (1.04) & E & (1.12) & E & (1.64) \\
& 3 & U   &    & D & (1.68) & S & (1.06) & S & (1.03) \\
& 4 &   &       &  U  &       &  D & (1.45) & S & (1.08) \\
& 5 &    &     &    U   &     & U   &         & D & (1.42) \\
& 6 &     &      &   U  &      & U   &        &  D & (2.73) \\
& 7 &      &      &     &     &    U   &   &      U & \\
& 8 & & & &  & U &  &U  &\\
& 9 & & && & U & &U &\\
&10 &&&& &&& U &\\
& 11 &&&&&&& U &\\
& 12 &&&&&&& U &\\ \hline

\end{tabular}
\end{center}
\end{table}


\subsection{Simulation Studies}
  We conduct a comprehensive study that evaluates the performance of
mTPI-2 and mTPI. 
Powered by crowd sourcing, we include a
study based on 1,774 scenarios and 6,013,460 simulated
trials, generated by 71 independent users of our existing tool,
NGDF \citep{yang2015integrated}. NGDF is a web tool that allows users to
design and simulate dose-finding trials based on various methods,
including 3+3, CRM, and mTPI. We take the scenarios and 
simulation settings (including sample size and number of simulated
trials per scenario) and simulate trials based on mTPI and
mTPI-2. Therefore, the scenarios we use are from NGDF users, which
constitute a crowd-sourcing exercise. Crowd sourcing typically allows
objective and unbiased assessment of various methods, since the
evaluators are a large   number   of different users, rather than
the inventors themselves.   

\begin{figure}[htbp]
\includegraphics[scale=0.3]{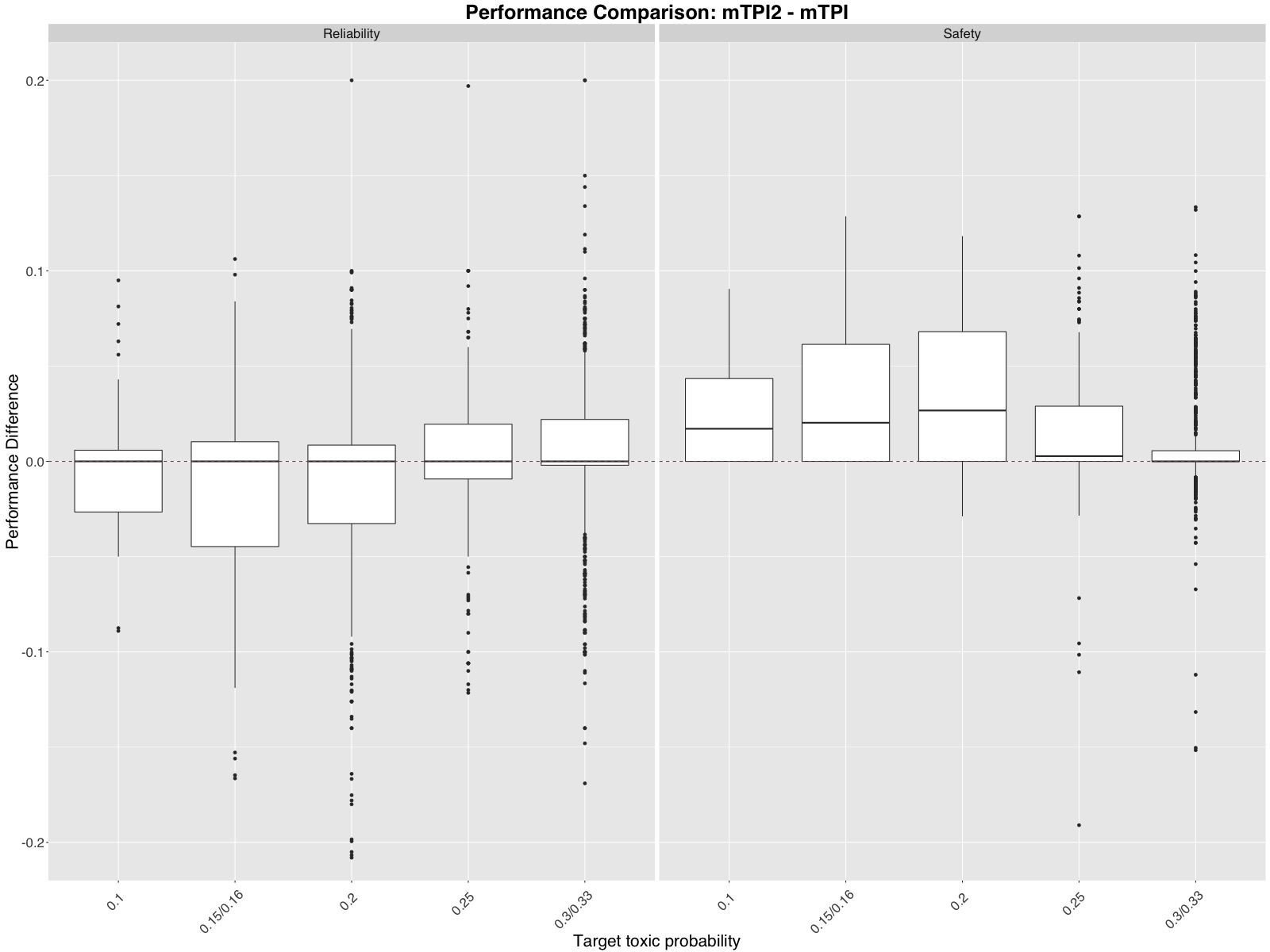}
\caption{Boxplots comparing the reliability and safety of mTPI and mTPI-2.} \label{fig:comp}
\end{figure}

We compare both methods in terms of reliability and safety, as
described in \citet{ji2013modified}. In particular, reliability is the
average percentage that the true MTD is
selected at the end of the trial, for a given scenario and across all
the simulated trials; and safety is the average
percentage of patients treated at or below the true MTD, for a given
scenario  and across all the simulated trials. So for each method, we
obtain 1,774 reliability values, one for each scenario. We then take
pair-wise differences between any two methods in their reliability
values for the same scenario, and plot the boxplots of the differences in the left half of
Figure \ref{fig:comp}. Each boxplot corresponds to a unique $p_T$ value of
the simulated trials. In the right half we show the boxplots for
safety comparisons in the same manner. 

Figure \ref{fig:comp} shows that when $p_T \leq 0.2$, mTPI is slightly
more reliable in identifying the true MTD than mTPI-2. However, when $p_T
> 0.2$, mTPI-2 is more reliable. What stands out is that mTPI-2 is
always safer than mTPI regardless of the $p_T$ values, which means that
mTPI-2 has less chance of assigning patients to overly toxic doses
than mTPI. In practice, mTPI-2 and mTPI are both easy to implement, only requiring
1) generating dose-assignment decision tables (e.g., in Figure
\ref{fig:tables}a) prior to trial initiation and 2) following
the decisions in the table during the course of the trial. 

\section{Software}
We have implemented mTPI-2 as an online tool at
\url{www.compgenome.org/NGDF}. It only requires a web browser, such as
Google Chrome, to access. The same website hosts mTPI, 3+3, and a version of CRM
which allows head-to-head comparison between mTPI-2 and these
designs. There is no need to download or maintain any software
package, and the web tool can be accessed anywhere via internet. In
our experience, the web tool runs successfully on a tablet such as
iPad or a smart phone such as iphone. This capability allows
investigators to use the design with great flexibility. A detailed
user manual is provided on the website to assist new users.

\section{Discussion}
We present mTPI-2, an improved mTPI design, to reduce the effect from the Ockham's razor
in the posterior inference. The mTPI-2 design is based on formal Bayesian decision theoretic framework,
adjusting for Ockham's razor. It mitigates some suboptimal decisions in
mTPI and provides theoretically optimal and intuitively sound decision
rules. As a result, mTPI-2   makes more refined actions that allow
more efficient exploration of different doses    in the dose finding
process.   

The mTPI-2 design hinges on 
user-provided quantities,   $p_T$,   $\epsilon_1$ and
$\epsilon_2$.   It treats   any
dose with toxicity probability smaller than $(p_T - \epsilon_1)$ or
larger than $(p_T + \epsilon_2)$ 
as being lower or higher than the MTD, respectively. Therefore,
these two values are the key input of the design and must   be   elicited from
physicians. For example, one can ask the physician what the highest
toxicity rate is that would still warrant a dose escalation ($p_T -
\epsilon_1$) and the
lowest rate ($p_T + \epsilon_2$) that would warrant a dose
de-escalation.   In this paper, we consider
$\epsilon_1=\epsilon_2$. Intuitively, when the two $\epsilon$'s are
not equal, the decisions can be altered in a nonsymmetric way such as
allowing more escalation than de-escalation or the opposite. This is an ongoing
research direction that we are currently pursuing. 


We focus on the comparison between mTPI and mTPI-2 in this paper. For
interested readers desired to compare mTPI-2 to the 3+3 design \citep{storer1989design}
 or the continual reassessment method (CRM, \citet{o1990continual}), we refer to \citet{ji2013modified} and \citet{yang2015integrated} who
compared mTPI to 3+3 and CRM through extensive simulation
studies, which serves as an indirect comparison to
mTPI-2. 

Innovatively, mTPI-2 is able to provide Bayes factors for each
decision so that investigators can assess the uncertainty behind it. These Bayes factors may provide
additional use for future work, such as allowing for randomization
between two different decisions when the value of Bayes factor
comparing the two decisions is very close to 1. 

The size of the equivalence interval serves as an ``effect size'' for
phase I dose-finding trials. This is an added benefit of
interval-based designs, such as mTPI and mTPI-2. A narrower
equivalence interval implies that the MTD must be identified with more
precision, and therefore demands a larger sample size. Also the sample
size will depend on the number of doses in the trial and the cohort
size, see \citep{ji2013modified} for a discussion. We intend to
address the sample size issue in a future work.

\clearpage

\clearpage
\section*{Appendix}

\subsection*{A. Proof of Theorem 1}
 Recall that $S(M_i)$ is the size of 
interval length 
for model $M_i$, $i \in \{E, S, D \}$. For example, for $M_E$, $S(M_E)
= p_T - \epsilon_1.$  

It suffices to show that the decisions rule $\cD_{\mbox{mTPI}}$
maximizes $E((1-\ell(i, M_j)) \mid \{x_d, n_d\})$, the posterior expected
utility, where utility is defined as one minus the 0-1 loss, i.e.,
$(1-\ell(i, M_j)).$ The posterior expected utility for action $i \in \{E,
S, D\},$ at dose $d$ is given by
\begin{eqnarray*}
  L(i, d) &=& \sum_{j \in \{E, S, D\}}  \ell(i, M_j) p \left(M_j \mid
    (x_d, n_d) \right)  \\
  & \propto & \sum_{j \in \{E, S, D\}} \ell(i, M_j) \int p(x_d \mid
  n_d, p_d) p(p_d \mid M_j) p(M_j) dp_d \\
  &=& \int p(x_d \mid n_d, p_d) p(p_d \mid M_i) p(M_i) d p_d \\
  &\propto& \int_{l_i}^{h_i} \frac{1}{S(M_i)} p_d^{x_d}
            (1-p_d)^{n_d - x_d} d p_d \\
  &\propto& \frac{Pr(M_i \mid \{x_d, n_d\})}{S(M_i)} \\
  &=& UPM(i, d)
\end{eqnarray*}
Therefore, the decision rule \eqref{eq:mTPI-rule} given by 
\begin{equation}
\cD_{\mbox{mTPI}} =  \arg \max_{i \in \{E, S, D\}} UPM(i, d) 
\end{equation}
maximizes the posterior expected utility, which is equivalent to
minimizing the posterior expected 0-1 loss.
\clearpage
\subsection*{B. Rate of Incomplete Beta Function}
We only need to consider the posterior probability of model $M_i$ in
the calculation of UPM, i.e., 
\begin{multline}
Pr(M_i \mid {x_d, n_d}) \propto \frac{1}{S(M_i)} \int_{l_i}^{h_i} p_d^{x_d}
            (1-p_d)^{n_d - x_d} d p_d  \\
\propto 
\frac{I_{h_i}(x_d + 1, n_d -x_d+1) - I_{l_i}(x_d +1, n_d - x_d
  +1)}{h_i - l_i} \label{eq:ibeta}
\end{multline}
where 
$$
I_{x}(p, q) = \frac{1}{B(p, q)} \int_0^x t^{p-1} (1-t)^{q-1} dt
$$
is the incomplete beta function, with 
$$
B(p, q) = \int_0^1 t^{p-1} (1-t)^{q-1} dt.
$$

Based on \citet{johnson2002continuous}, 
$$
I_x(p, q) \approx \Phi(z)
$$
where
\begin{multline}
z=\frac{k}{|q-0.5-n(1-x)|} \left\{\frac{2}{1+(6n)^{-1}} \left[\left(q-
        0.5  \right)\log \left\{\frac{q-0.5}{n(1-x)}  \right\}
      + (p-0.5) \log\left\{\frac{p-0.5}{nx}  \right\}  \right]  \right\}^{1/2},
\end{multline}
$n= n_d -1$ and $k=n_d - x_d - 1/3 - (n_d + 1/3)(1-x)$.  
When $x_d=3$ and $n_d =6$, the incomplete beta function can be shown
to be approximated by $$\Phi(sgn(x-0.5)) *
\sqrt{-7\log(x(1-x))}).$$ Based on Feller (1968), this can be
approximated by 
$$
 I(x>0.5)*\frac{1}{2} + sgn(x-0.5)* \frac{e^{-y^2/2}}{\sqrt{2 \pi} y},
 \quad y =  \sqrt{-7\log(x(1-x))}, 
$$
which equals
$$
I(x>0.5)*\frac{1}{2} + sgn(x-0.5)*\frac{\{x(1-x)\}^{1/14}}{\sqrt{-14 \pi \log(x(1-x))}}. 
$$
A numerical evaluation reveals that when $x$ takes values at 0.25,
0.35, and near 1, the expression of \eqref{eq:ibeta} favors model $M_D$
in which $h_i = 1$ and $l_i=0.35$ over model $M_S$ in which $h_i =
0.35$ and $l_i = 0.25$. Unfortunately, there is no general conclusion
on the value of \eqref{eq:ibeta} for any $x_d$ and $n_d$ values, which
makes the theoretical derivation difficult. The above derivation
pushes forward the theoretical development for the incomplete beta
function  in that it gives the ratio of
$(x(1-x))^{1/14}/\sqrt{-log(x(1-x))}$. However, the entire function if
not monotone with a mode at 0.5, which makes it difficult to evaluate
the magnitude of \eqref{eq:ibeta} as a difference of two incomplete
beta functions. It is known that the
analytic expression of incomplete beta function is still an open
research question \citep{swaminathan2007convexity}. Therefore, we leave the further
theoretical development to future work. 

\end{document}